\documentclass[a4paper,10pt]{article}

\usepackage{natbib}
\usepackage[koi8-r]{inputenc}
\usepackage{aas_macros}
\usepackage[russian,english]{babel}
\usepackage{graphicx}

\def\d{^{d}\kern-2.1mm .\kern+.6mm}
\def\apgt{\ {\raise-.5ex\hbox{$\buildrel>\over\sim$}}\ }
\def\aplt{\ {\raise-.5ex\hbox{$\buildrel<\over\sim$}}\ }
\def\m{^m\kern-7pt .\kern+3.5pt}
\newcommand{\ms}{\mbox {$M_{\odot}$}}
\newcommand{\rs}{\mbox {$R_{\odot}$}}

\usepackage{amsmath}
\usepackage{amssymb}
\usepackage{rotating}
\usepackage[T2A]{fontenc}
\begin{document}

\righthyphenmin=2
\begin{center}
{\bf A MODEL OF THE POPULATION OF HELIUM STARS \\
IN THE GALAXY I. LOW-MASS STARS}

\vskip 0.3cm

L.R.~Yungelson and A.V.~Tutukov
\end{center}

\vskip 0.2cm
\begin{center}
\textit{Institute of Astronomy of Russian Academy of Sciences} 

\textit{lry, atutukov@inasan.ru}

\end{center}

{\bf Abstract}
By means of population synthesis we model the Galactic ensemble of helium stars. It is assumed that all helium stars are formed in binaries. Under this assumption, single helium stars are produced by the mergers of helium remnants of components of close binaries (mainly, by merging helium white dwarfs) and by disruption of binaries with helium components in supernovae explosions. The estimate of the total birthrate of helium stars in the Galaxy is $0.043$\,yr$^{-1}$, their total number is estimated as  $4 \times 10^6$. The rate of binarity in the total sample is 76\%.
We construct a subsample of low-mass ($M_{\rm He} \lesssim 2\,\ms$)
 helium stars limited by observational selection effects:  stellar magnitude ($V_{\rm He} \leq 16$), ratio of stellar magnitudes of components in binaries   ($V_{\rm He}\leq V_{\rm comp}$), lower limit of the semiamplitude of radial velocity that is necessary for discovery of binarity ($K_{min}=30$\,km/s).
The parameters of this ``observable'' sample are in satisfactory agreement with the parameters of the observed ensemble of sdB stars. In particular, in the selection-limited sample binarity rate is 58\%. We analyze the relations between orbital periods and masses of helium stars and their companions in systems with different combinations of components. We expect that  overwhelming majority ($\sim 90$\%) of unobserved components in binary sdB stars  are
white dwarfs, predominantly, carbon-oxygen ones.




\section{Introduction}
\label{sec:intro}

The aim of the present paper is investigation of nondegenerate helium stars.
Among observed objects they are represented by two groups -- helium subdwarfs
and Wolf-Rayet stars (WR). The interest to helium stars is enhanced by the fact
that they are related to the stars of another groups: AM CVn stars, R CrB
stars, double-degenerates (possible progenitors of type Ia supernovae). The
most massive helium stars are progenitors of type Ib and Ic supernovae. The
optical component of one of the most remarkable X-ray sources -- Cyg X-3 is a
helium star. Among stars with spectroscopically determined masses the most
massive is the close binary WR~20a; both components of this system are helium
stars  with $M \sim 80$\,\ms~\citep{rauw+04,rauw+05}.

Helium subdwarfs are subdivided into three groups: sdB stars that have
hydrogen-rich atmospheres, sdO stars with atmospheres dominated by helium, and
an intermediate group of sdOB stars. The masses of helium subdwarfs that are
usually derived from their location in the ``effective temperature  $\log
T_{\rm eff}$ -- effective gravity $\log g$'' diagram, are close to
0.5\,$M_\odot$~ \citep{heber86,saffer+94}.

These estimates are consistent with masses determined for components of
eclipsing systems that range from 0.48\,$M_\odot$  to 0.54\,$M_\odot$
(see~\citet{drechsel01} and references therein) and with masses obtained from
asteroseismological data: $0.49\pm0.02\,M_\odot$
(PG0014+067~\citep{brassard01}), 0.47\,$M_\odot$ (Feige~48~\citep{reed04}). One
has to have in mind, however, that the evolutionary tracks of numerous low-mass
($M_{\rm wd} \lesssim 0.3\,M_\odot$) helium white dwarfs in the stage of
contraction of their hydrogen envelopes, also cross the region of the  $\log
T_{\rm eff}$ -- $\log g$ diagram populated by helium subdwarfs. Respectively,
some of the stars that are classified as subdwarfs may happen to be not helium
stars but relatively hot helium white dwarfs. For instance, the mass of
HD188112 that is classified as sdB is 0.24\,$M_\odot$  only, i. e., it is a
precursor of a helium white dwarf~\citep{heln03}. As another example of
uncertainty one may consider HS2333+3927: position of this star in the  $\log
T_{\rm eff}$ -- $\log g$ diagram ``satisfies'' both tracks of a 0.29\,$M_\odot$
white dwarf and of a  0.47\,$M_\odot$ helium star with a thin hydrogen
envelope~\citep{heber_hs2333}.

The lower limit to the range of masses of WR stars is  $\simeq 7\,M_\odot$~
\citep{nl00}. However, according to the modern theory of the evolution of close
binary stars, mass spectrum of helium stars has to be continuous. It is
possible that the WR phenomenon is observed for stars with  mass $\gtrsim
7\,M_\odot$ only, because for mass-luminosity relation typical for the lower
mass stars radial pulsations that may result in generation of shock waves that
produce  dense enough  stellar wind are not excited (see, e.
g.,~\citet{fadeyev03a}).

There are several channels for production of helium stars. Stars with ZAMS
masses exceeding  $\sim 50\,M_\odot$ turn into WR stars due to stellar wind
mass loss~\citep{conti79}. It is possible that single stars and components of
wide binaries of moderate mass may almost completely lose their hydrogen
envelopes immediately before helium flash or during the
latter~\citep{cox_salp61,dcruz96}. In close binaries, stars with helium cores
and thin hydrogen envelopes form due to the mass loss after Roche lobe
overflow~\citep{kw67,pac67b}.  Single helium stars may form due to the merger of
helium white dwarfs or nondegenerate helium stars in close
binaries~\citep{ty90,iben90,sj00}.

The main scenarios for formation of low-mass close binaries were considered
by~\citet{it85,ty90,iben90,hpmmi02,hpmm03}. In particular, it has been shown
that the location of helium remnants of components of close binaries and  of
the products of their merger and their evolutionary tracks in the
Hertzsprung-Russell diagram and in the $\log T_{\rm eff} - \log g$ diagram well
fit observations~\citep{iben90,hpmmi02,hpmm03}. Han et al.
have considered in detail the relations between different formation channels
and have shown that, if selection effects are taken into account, it is
possible to explain the main parameters of observed helium subdwarfs.

In the present paper, by means of population synthesis,  we construct a model
of the ensemble of Galactic helium stars  and analyze the properties of its
low-mass subsystem (i.e., of helium subdwarfs). We apply the population
synthesis code used by \citet{ty02}. At difference to~\citet{hpmmi02,hpmm03}, we compare
with observations a subset of model stars limited by stellar magnitude, since
such a limit exists for the main catalog of hot subdwarfs
PG~\citep{palomar_green}. We pay main attention to some aspects of binarity of
hot subdwarfs, in particular, to the distribution of the pairs of subdwarfs
with different companions over orbital periods.

Massive helium stars and the products of their evolution are considered in the
next paper of the series (Paper II).

											In
Sect. \ref{sec:popsyn} we describe the basic assumptions used in the population
synthesis code that are immediately relevant to the modeling of formation of
helium stars. The main results are presented in Sect. \ref{sec:results}. 
Discussion follows in Sect.
\ref{sec:concl}.                                                                                                                   
\section{Population synthesis for binary stars} \label{sec:popsyn}

Both high- and low-mass  helium stars are formed when one of the components of
a close binary system overflows its Roche lobe in the hydrogen shell burning
stage~\citep{kw67,pac67b} (in the co called ``case B'' of mass exchange). In the
population synthesis code we have the following assumptions relevant to the
mass exchange in case B.

Components of close binaries with ZAMS masses $M_{\rm i} <  2.8\,M_\odot$~form
degenerate helium dwarfs.  Components of close binaries that initially had mass
higher than  2.8\,$M_\odot$ (or accumulated that much mass as a result of mass
exchange) produce nondegenerate helium stars. The mass of the latter is 
related to $M_{\rm i}$ by the following approximation to the results of
evolutionary computations~\citep{ty73a,it85}:
\begin{gather}
 \label{eq:mimf}
  M_{\rm He}/M_\odot = \max(0.066(M_{\rm i}/M_\odot)^{1.54}, 0.082(M_{\rm
i}/M_\odot)^{1.4}).
\end{gather}
Equation (\ref{eq:mimf}) gives the minimum
mass of helium stars close to 0.35\,$M_\odot$, in a reasonable agreement with
results of more detailed evolutionary computations~\citep{hte00}. We neglect the
dependence of the mass of helium remnant on the instance of the Roche lobe
overflow that is not very significant in the case B of mass exchange.

Based on the results of evolutionary computations~\citep{ty73,it85}, we
approximated the lifetime of stars in the region of Hertzsprung-Russell diagram
populated by helium stars as 
\begin{gather}
 \label{eq:the}
  \log T_{\rm He} =
   \begin{cases}
   7.15 - 3.7\log(M_{\rm He}/M_\odot) & \text{for $M_{\rm He} \leq 1\,M_\odot$}, \\   
   7.15 - 3.7\log(M_{\rm He}/M_\odot)+2.23[\log(M_{\rm He}/M_\odot)]^{1.37} & \text{for $M_{\rm He}    > 1\,M_\odot$}.
   \end{cases}
\end{gather}

Stellar wind mass loss in the main-sequence stage is described by initial-final
mass relations~\citep{vanbev91}:
\begin{gather}
 \label{eq:hsw}
  M_{\rm f} =
   \begin{cases}
    M_{\rm i} - 0.00128 M_{\rm i}^{2.5} & \text{for $M_{\rm i} \leq 30\,M_\odot$}, \\   
    1.77 M_{\rm i}^{0.763} & \text{for $M_{\rm i} > 30\,M_\odot$}.
   \end{cases}
\end{gather}

It is assumed that in the core-hydrogen burning stage the stars with ZAMS mass
in excess of 50\,$M_\odot$ lose by stellar wind so much matter that they become
helium stars immediately after completion of main-sequence stage and never fill
Roche lobes. Thus, only helium stars with mass  $\gtrsim  27$\,$M_\odot$\ may
be products of the evolution of single stars or of components of wide binaries.

Mass loss by helium stars with $M_{\rm He} \geq 5\,M_\odot$\ is described by
the following approximation to the observed mass-loss rates by WR-stars (see
compilation by \citet{nl00}):
\begin{gather}
 \label{eq:hesw}
  \dot M_{\rm He} =
\min(1.38 \times 10^{-8}(M_{\rm He}/\ms)^{2.87},~10^{-4}) M_\odot/\textrm{yr}.
\end{gather}
The analytical expression in Eq. (\ref{eq:hesw}) is suggested by~\citet{nelemans01}. The slope of the $\dot M_{\rm
He}(M_{\rm He})$ relation changes at $M_{\rm He} \approx 22.1\,M_\odot$.

Some signs of the existence of stellar wind are suspected for several hot
subdwarfs only~\citep{hmmkd03};  applying the model of radiative stellar wind,
\citet{vink04} estimated the rate of mass loss as $\dot{M} \sim
10^{-11}\,M_\odot$/yr. Taking into account the lifetime of helium stars one may
claim that the only effect that such a mass loss may have upon evolution of hot
subdwarfs is the loss of the remainders of hydrogen envelope and, possibly,
change of the chemical composition of the subdwarf atmosphere.

Helium stars with initial mass $M_{\rm i} = 2.8 \div 5\,M_\odot$\ do not expand
after helium exhaustion in their cores and become carbon-oxygen white dwarfs
remaining compact. If $M_{\rm i}=(5 - 11.6)\,M_\odot$, helium stars finish
their evolution as carbon-oxygen white dwarfs after expansion in the
helium-shell burning stage and repeated  Roche lobe overflow that is
accompanied by an insignificant mass loss~\citep{it85}. 

Stars with ZAMS mass  $M_{\rm i} = 11.6 \div 30\,M_\odot$\ produce neutron
stars with  $M_{\rm NS}=1.4\,M_\odot$. Initially more massive stars end their
lives as black holes. Formation of black holes is accompanied by the loss of 20\%
of the presupernova mass. It is assumed that supernovae explosions are 
spherically symmetric, i. e., nascent neutron stars and black holes do not get
kicks (see  detailed discussion in Paper II).

Stars with $M_{\rm i} \leq 11.6\,M_\odot$\ that overflow Roche lobe in the AGB
stage (``case C'' of mass exchange) produce white dwarfs; the mass of the
latter depends on the instance of Roche lobe overflow. More massive stars
produce neutron stars or black holes, like in case B of mass exchange.

\begin{figure}[t!]
\includegraphics[scale=0.7,angle=-90]{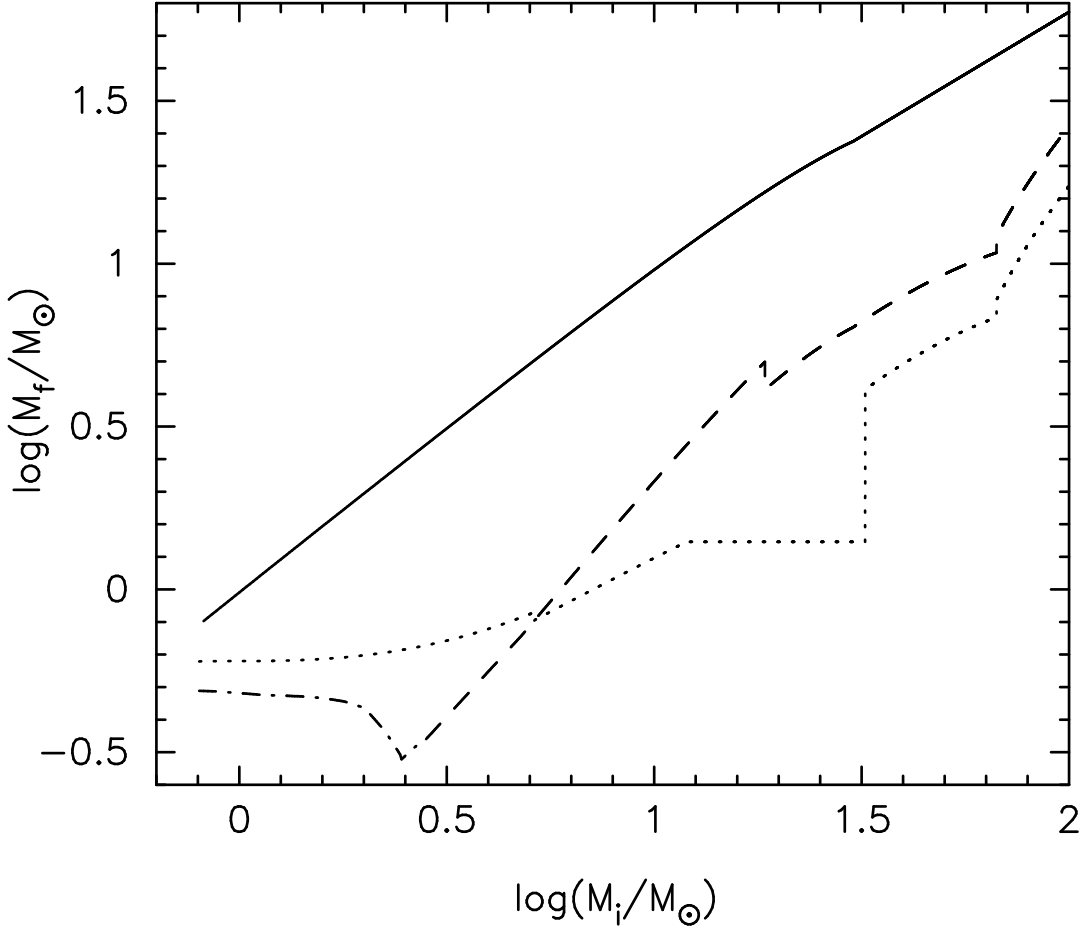}
\caption{Relation between ZAMS mass of stars $M_i$, their mass at the end of
main-sequence stage (solid curve), mass of helium stars produced by them
(dashed curve), and mass of the final products of evolution: helium white
dwarfs (dot-dashed curve), carbon-oxygen, oxygen-neon white dwarfs, neutron
stars, and black holes (dotted curve). For stars with $M_i \lesssim 5$\,\ms\
the upper limit of white dwarf mass attained in case C of mass exchange is
shown. }
\label{fig:mi_mf}
\end{figure}

The relation between initial mass of stars, their mass at the end of
main-sequence stage, mass of helium stars produced by them, and mass of the
final products of evolution is plotted in Fig.~\ref{fig:mi_mf}.                  

We have applied the following assumptions on the character of mass-exchange
process, based on results of numerical modeling~\citep{tfy82}.
If both donor and accretor are main-sequence stars, the donor has radiative envelope, and mass ratio of components $q=M_{\rm a}/M_{\rm d} \geq 0.8$,
mass exchange is conservative and proceeds in the time scale of the nuclear evolution of the donor. If  $0.5 \leq q< 0.8$, the donor loses mass in the thermal time scale $t_{th1}$, but accretion by the main-sequence secondary is limited by the rate $\dot{M}_{th2}$, that corresponds to the thermal time scale of the latter. The mass that may be accreted by the companion is limited by   $\Delta M = \dot{M}_{th2}t_{th1}$. Variation of the separation of components $a$
is calculated under assumption that  $\Delta M$ is accreted conservatively, but the rest of the donor envelope with mass $\Delta M_t=M_{\rm i} - M_{\rm f}-\Delta M$ is lost from the system; the latter process is described by the equation
\begin{gather}
\label{eq:j}
\frac {\Delta J}{J} = \gamma \frac{\Delta M_t}{M_t},
\end{gather}
where $J$ is angular momentum of the binary, $M_t=M_1+M_2$, $\gamma=1.5$.
The validity of Eq.  (\ref{eq:j}) with the parameter  $\gamma=1.5$ is based on
the results of reconstruction of the mass-exchange history in  observed close
binaries containing white dwarfs and helium
subdwarfs~\citep{nvy+00,nelemans_tout05}.

When the mass ratio of components $q < 0.5$ or donor has convective envelope
extending down to more than 50\% of stellar radius, mass loss occurs in
dynamical time scale and a common envelope is formed. Variation of component's
separation is computed based on the balance of the orbital energy of the binary
and donor envelope binding energy~\citep{ty79}:
\begin{gather}     
\label{eq:ce}
\frac{(M_{\rm i} +m) \; (M_{\rm i} - M_{\rm f})}{ a_{\rm i}}  
 = \alpha_{\rm ce} \; \left[ \frac{M_{\rm f}\; m}{a_{\rm f}} -
\frac{M_{\rm i} \; m}{a_{\rm i}} \right].
\end{gather}
In the Eq. (\ref{eq:ce}) indexes  ${\rm i}$ and   ${\rm f}$ are related to the
initial and final mass and separation of components, $m$ is accretor mass,
$\alpha_{\rm ce}$ is so-called ``common envelope parameter'' that describes the
efficiency of deposition of released energy into common envelope. Trial
computations have shown that the best agreement with observations is obtained
for $\alpha_{\rm ce}=2$, irrespective to the evolutionary state of the donor. 

If accretor is a white dwarf, neutron star or a black hole, Eq. (\ref{eq:ce}) is always applied.

Nondegenerate single helium stars may be products of the  evolution of close
binaries. We consider three channels for the formation of single helium stars:
the merger of two helium white dwarfs or of a helium and carbon-oxygen white
dwarfs and the merger of a helium white dwarf and a nondegenerate helium star. In the
first two cases the merger is a result of the loss of systemic angular momentum
via gravitational waves radiation (GWR). Helium star and a helium white dwarf may
merge either due to momentum loss via GWR or due to the loss of momentum in a common
envelope that forms after helium exhaustion in the core of nondegenerate star and its expansion
(for $M_{\rm He} \approx (0.8 - 2.8)\,M_\odot$).~If $M_{\rm d}/M_{\rm a} \gtrsim 2/3$,
the donor is disrupted in dynamical time scale and forms, most probably, a disk
around its companion~\citep{pw75,ty79a}.   Accretion in such systems is not studied
as yet, but one may expect that the rate of accretion from the disk is defined,
primarily, by the viscosity of the disk matter. The total minimum mass
$M_{\rm He}^{min}$ for formation of a nondegenerate helium star by the merger of white
dwarfs is,  essentially, still a free parameter. It is clear only that it has
to exceed the minimum mass of nondegenerate helium stars  $\sim 0.3\,M_\odot$\
(see, for instance,~\citet{iben90}). We assumed $M_{\rm He}^{min}=0.4\,M_\odot$. The
latter value is in a good agreement with results of
computations  \citep{iben90,hpmmi02}, see also discussion in Sect.
\ref{sec:results}. The lifetime of merger products of helium white dwarfs or
of helium stars and  helium dwarfs was also calculated by means of Eq.
(\ref{eq:the}), though, the latter may somewhat overestimate the lifetime of
merger products as helium stars (the overestimate is $\lesssim 20$\%,
depending on the mass of the merger product,  see Eq. (1) in~\citet{iben90}.
For the  merger products  of helium and carbon-oxygen dwarfs, their lifetime is
defined by ``competition'' of two processes: helium burning in the shell and
stellar wind mass loss. We have to note that the merger of helium dwarfs
provides in our model formation of about 90\% of single helium stars.

Based on the studies of eclipsing, spectroscopic, and visual
binaries~\citep{kpty81,pty82,kpty85,vtykp88,kpty89}, we assumed that all
stars are born in the
binaries with orbital separations $a$ ranging from
$6(M_1/M_\odot)^{1/3}$\rs\ to $10^6$\,\rs. Birthrate function for binaries was
taken as~\citep{pty82}:
\begin{gather}  
\label{eq:brate}
\frac{dN}{dt} = 0.2 d (\log a) \frac{dM_1}{M_1^{2.5}} f(q) dq,   
\end{gather}
where  $M_1$ is the mass of initially more massive component of the system, $f(q)$ is the distribution of binaries over mass ratios of components $q=M_2/M_1$ normalized to 1. We have assumed that for close binaries $f(q)=1$.
Normalization of function (\ref{eq:brate}) assumes that one binary system with $M_1 \geq 0.8\,M_\odot$ is born annually in the Galaxy. According to (\ref{eq:brate}), about 40\% of all binaries are ``close''. Star formation rate was set constant for 13.5 Gyr. Evidently, this assumption does not influence the results, since the lifetime of even least massive helium stars does not exceed 
$\sim$\,1Gyr.

Our model assumes that all stars are formed in binaries. Accordingly, we did
not consider the possibility of formation of low-mass helium stars as an
outcome of evolution of single stars or of the components of wide binaries
(see, e. g.,~\citet{dcruz96}). The  authors of
the latter study assume that precursors of
helium stars are red giants that lost almost whole hydrogen envelope
immediately before helium flash or in the course of the flash. The mechanism of
mass loss is unknown and the rate of mass loss is considered as a free
parameter.  The real rate of binarity of hot  subdwarfs is difficult to
determine, since for usually applied observational methods selection effects
preclude detection of binary subdwarfs with orbital periods longer than 200 to
300 day~\citep{mhmn01,nap_ehb04}.     It is evident that \textit{a priori} one
cannot exclude the possibility of descent of low-mass helium stars from single stars or components of wide binaries, but, as we shall show below, it is quite possible to explain
formation of sufficiently large number of single helium stars by  the mergers
of helium white dwarfs and/or nondegenerate helium stars. 

\section{MAIN RESULTS}
\label{sec:results}

\subsection{General properties of the population of low-mass helium stars}
\label{sec:general}

The main aim of our study is the modeling of the ensemble of helium stars and
of their distributions over main parameters that, potentially, can  be derived
from observations -- masses and orbital periods of binaries, as well as the 
analysis of the relations between parameters of close binaries with helium
components.   

As it was noted in Sect.  \ref{sec:popsyn}, helium stars in binaries form as a
result of Roche lobe overflow by stars that are in the hydrogen-shell burning
stage. These stars may be either primary or secondary components of the system. In
the first case the companions to the nascent helium stars are main-sequence stars, while in the second case they  are white dwarfs, neutron stars or
black holes or, on very rare occasions -- also helium stars.

In the Table we list total birthrate and current
number of helium stars in the Galaxy, found if assumptions described in Sect.
\ref{sec:popsyn} are applied. The dominant constituents of the population of
helium stars are binaries with companions -- main-sequence stars and with
carbon-oxygen white dwarfs. The number of the objects of the  third main group 
-- single stars -- depends on the assumptions on the minimum mass necessary for He-ignition
in the merger product of two helium white dwarfs: if this critical
mass is reduced from $M_{\rm He}^{min} =0.4\,M_\odot$\ to  0.35\,$M_\odot$, the
number of single helium stars increases by approximately 8\%. About 0.8\% of
all helium stars have neutron star companions. The number of such binaries may
be overestimated by an order of magnitude (see, e.
g.,~\citet{lpp96b,lpp97b,py98,lyhnp05}), since in the present model we do not
take into account the possibility  that  a nascent neutron star may get at
birth  a kick which may result in disruption of the system (see more detailed
discussion in Paper II).

\renewcommand{\tabcolsep}{10pt}%
\begin{table}[t!]
\label{tab:birate}
\flushleft
\caption{Birthrate and number of helium stars.}
\bigskip
\begin{tabular}{lccc}
\hline
Type of star & Birthrate  & Total  & Subsample  \\
 & per yr & number  & with $V_{\rm He} \leq 16$ \\[1mm]
\hline 
Single             & $0.67\times 10^{-2}$ & $0.96\times 10^6$ &  $0.64\times 10^4$ \\
Main-sequence companion& $0.21\times 10^{-1}$ & $0.13\times 10^7$ &  $0.24\times 10^3$ \\
He white-dwarf companion & $0.50\times 10^{-2}$ & $0.29\times 10^6$ &  $0.18\times 10^4$ \\
CO white-dwarf companion & $0.66\times 10^{-2}$ & $0.13\times10^7$  &  $0.64\times 10^4$ \\
ONe white-dwarf companion & $0.80\times 10^{-3}$ & $0.31\times10^5$  &  $0.15\times 10^3$ \\
Neutron star companion& $0.19\times 10^{-2}$ & $0.31\times10^5$  &  $0.21\times 10^3$ \\
Black hole companion & $0.14\times 10^{-3}$ & $0.96\times10^4$  &  $0.66\times 10^2$ \\
Binary helium star & $0.34\times10^{-3}$ & $0.61\times10^4$ &   $0.58\times10^2$ \\[1mm]
\hline
\end{tabular}
\end{table}

\begin{figure}[t!]
\includegraphics[scale=0.7,angle=-90]{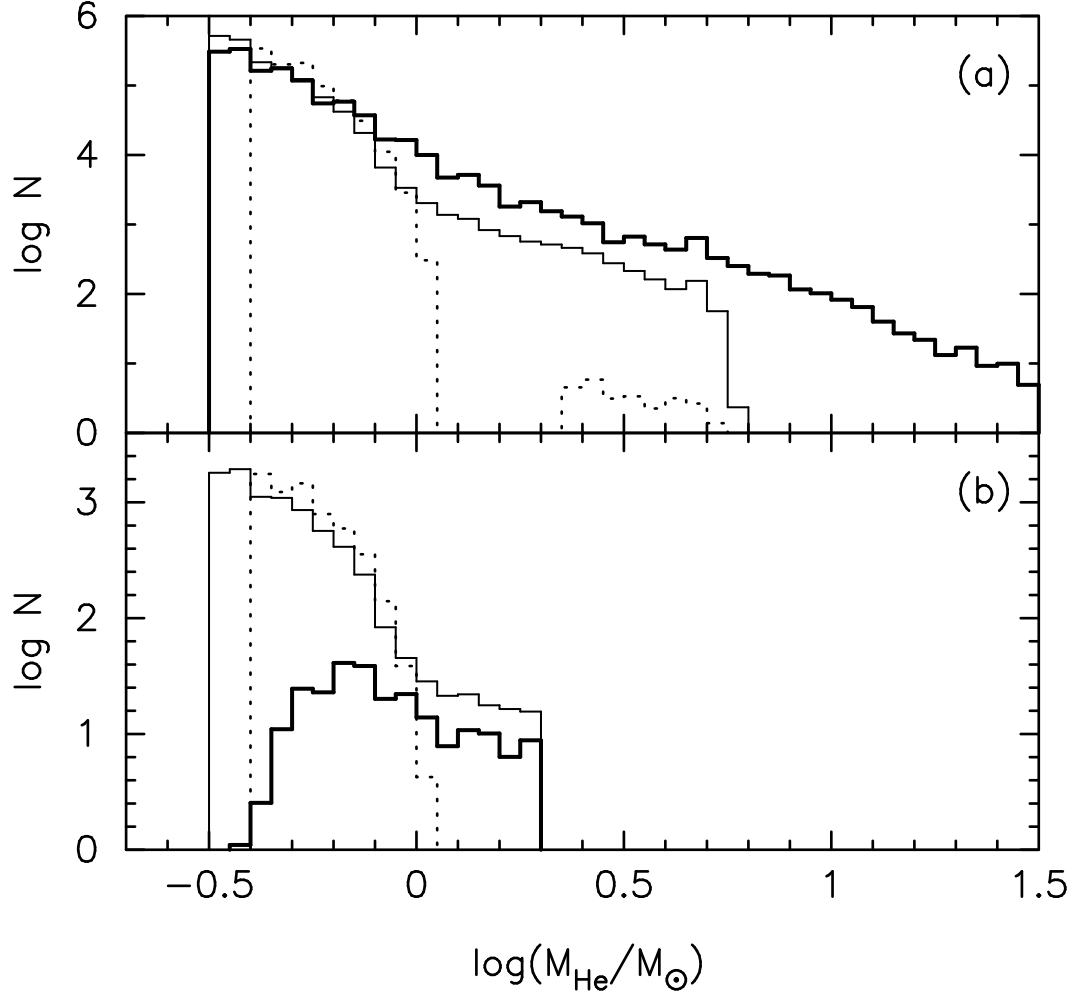}
\caption{Distribution of helium stars over mass in the total sample (a) and in the sample
limited by selection effects (b). Thick solid line shows helium stars with
main-sequence companions, thin solid line -- stars with white-dwarf companions,
dotted line -- single helium stars. 
}
\label{fig:wmall}
\end{figure}

\newpage
\begin{figure}[!t]
\includegraphics[scale=0.8]{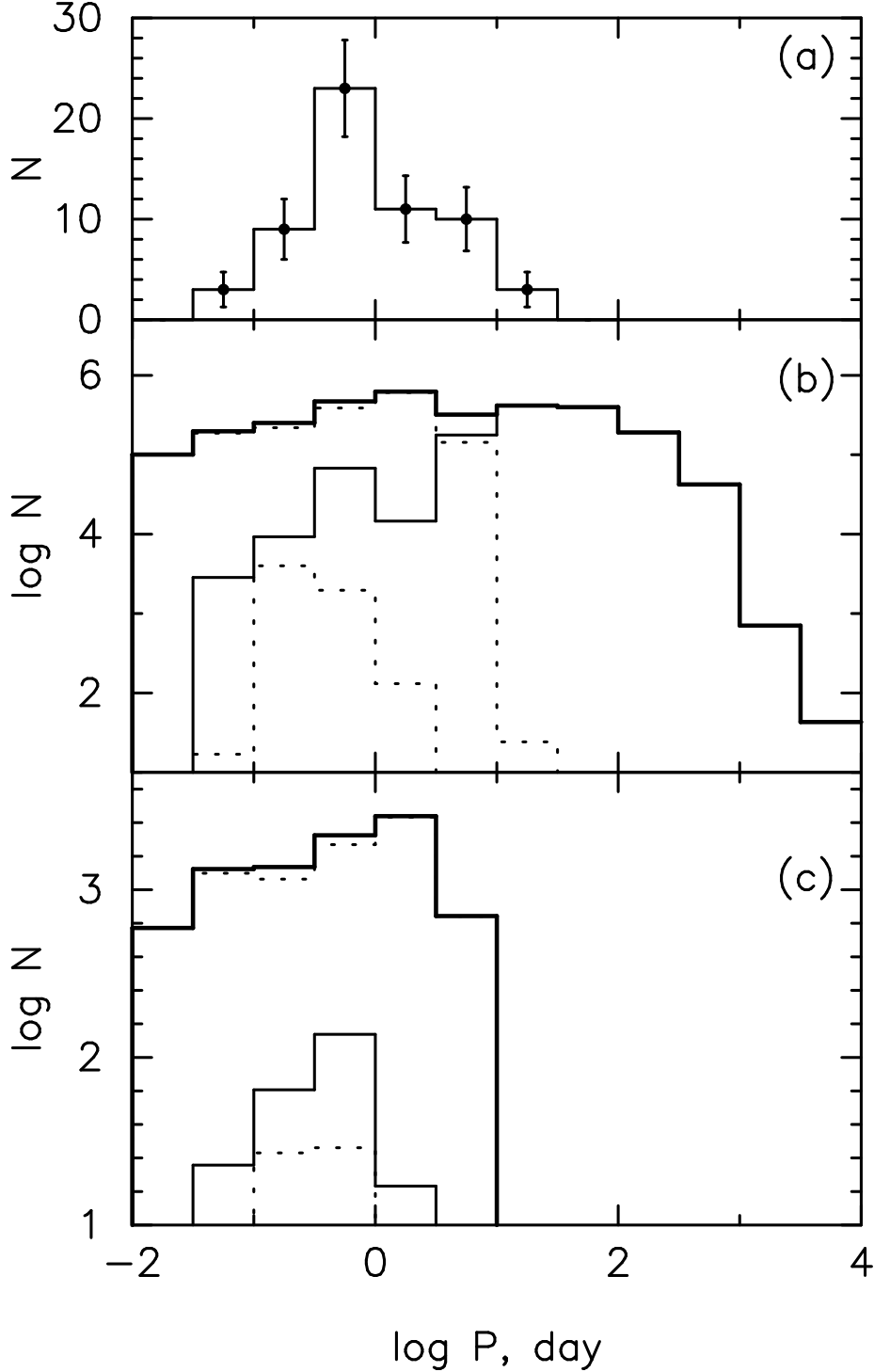}
\caption{Distribution of helium stars over orbital periods. (a) -- observed subdwarfs
(L. Morales-Rueda, priv. comm.), (b) -- model sample for total population.
Upper thick line shows the summary distribution, thin solid, upper and lower
dashed lines show the input of stars with main-sequence, white-dwarf companions
and helium secondaries into total population. (c) -- distribution over periods
in the sample limited by selection effects; the meaning of the lines is like in
panel (b). }
\label{fig:wpsel+obs}
\end{figure}

Total model Galactic birthrate of helium stars is $0.043$\,yr$^{-1}$, total
number of helium stars is $4 \times 10^6$. The rate of binarity of helium stars
is 76\%. Since progenitors of helium stars descend from main-sequence stars of
moderate and high mass, the semi-thickness of the Galactic disk for them may be
taken equal to 200 pc (this agrees with empirical $z$-scale for helium
subdwarfs~\citep{heber86,downes86,boer+95}; then the volume of the disk component
of the Galaxy is equal to $2.8 \times 10^{11}$\,pc$^3$, while spatial density
of helium stars is equal to  $1.4 \times 10^{-5}$\,pc$^{-3}$. The latter
estimate is considerably higher than the observational estimates: $2 \times
10^{-6}$\,pc$^{-3}$~\citep{heber86}, $4 \times
10^{-6}$\,pc$^{-3}$~\citep{downes86}. However, as we shall show further, a
considerable fraction of low-mass helium stars may be ``lost'' in observations
due to selection effects.

In Fig.~\ref{fig:wmall} we show the distribution of helium stars with
different companions over mass (for the total population and for the subset
limited by selection effects, see below). In all groups the mass of helium
stars strongly concentrates to the minimum value ($\simeq 0.4\,M_\odot$). The
main group  of single objects consists of merger products of white dwarfs and
white dwarfs and helium stars; very scarce single stars with mass from 2.2 to
5.6\,$M_\odot$\ are merger products of helium stars and helium dwarfs or former
companions of binaries that were disrupted by supernovae explosions. 
In the systems with companions -- helium dwarfs, the
mass of helium stars is limited by $\simeq 0.9\,M_\odot$. This maximum
corresponds to the case of completely conservative first mass exchange and
totally nonconservative second mass exchange, when Roche lobe is filled by
initially less massive component of the system, i. e., to the following
sequence of combinations of masses of companions: $(M_1,~M_2) = (2.8\,\ms,
2.8\,\ms) \rightarrow (0.4\,\ms,  5.2\,\ms) \rightarrow (0.4\,\ms, 0.9\,\ms)$.
The upper limit of helium-star mass in the systems with CO- or ONe-companions
is related to a similar scenario and to the upper limit to the mass of
progenitors of white dwarfs:  $(M_1,~M_2) = (11.4\,\ms, 11.4\,\ms) \rightarrow
(1.35\,\ms,  21.45\,\ms) \rightarrow (1.35\,\ms, 5.7\,\ms)$. In the latter
systems companion to the white dwarf explodes in the course of further
evolution and disrupts the system.

\begin{figure}[!t]
\includegraphics[scale=0.7,angle=-90]{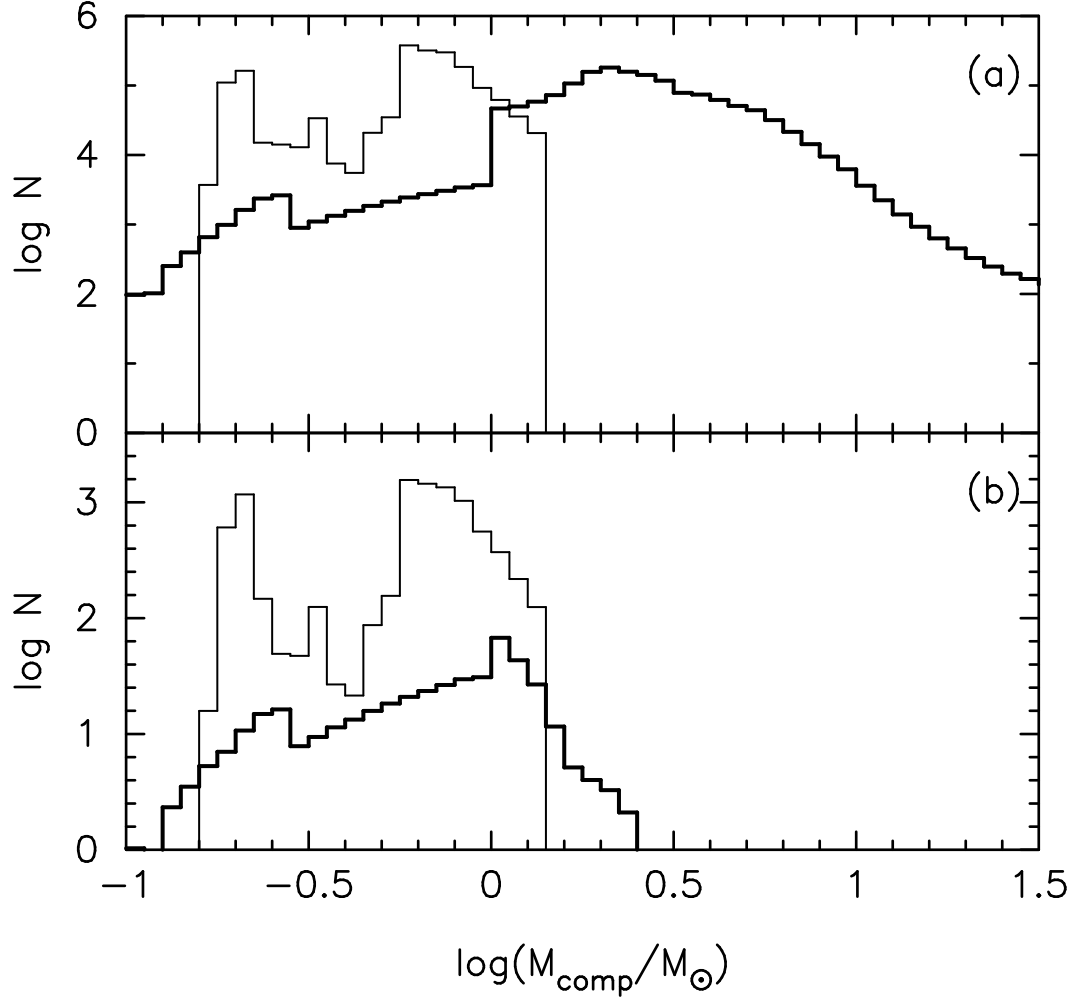}
\caption{Distribution of companions to helium stars over mass. (a) -- total model
sample. Thick line -- distribution of main-sequence companions, thin line --
distribution of white-dwarf companions. (b) -- model sample limited by
selection effects; the meaning of the lines is like in panel (b). }
\label{fig:wcall}
\end{figure}

In Fig. \ref{fig:wpsel+obs}b we show the distribution of helium stars over
orbital periods.  
The lower limit of orbital periods
is determined by condition for the merger of components in the common envelope.
For the systems with companions -- main-sequence stars the upper limit of the
periods is the maximum period that can be attained in the process of conservative
mass exchange. For the systems with white-dwarf companions the upper limit of the
periods is  determined by reduction of orbital separations in the common
envelopes. 
There are systems with $P_{orb} \sim 0.01$\,day in the model sample. The shortest orbital period of the known observed systems is $P=0.073$\,day
(PG1017-086). Existence of model systems with shorter orbital periods may be
due to the fact that, in the population synthesis code in the formal procedure
that checks whether the merger of components in the common envelope is
possible, the radius of a helium star with mass equal to the mass of the helium
core of the donor is used, while a more realistic procedure has to employ much
larger radius of the inert helium core itself. It is also possible that the common
envelope parameter $\alpha_{\rm ce}$ is somewhat larger than we assume. Another
reason may be underestimate of the radii of helium stars that depend on the
mass of hydrogen envelopes. Thus, the shortest orbital period systems shown in
Fig.~\ref{fig:wpsel+obs}  have, probably, to be classified as single helium
stars. Then their number has to be lower than shown in the Figure, because of
shorter lifetime of merger products of helium white dwarfs and helium
stars.       

The distribution of the companions to helium stars over mass is shown in
Fig.~\ref{fig:wcall}a. Two discontinuities in the distribution for stars with
main-sequence companions, occurring close to 0.3\ms\ and 1.0\,\ms, are due to
the loss of mass and angular momentum by main-sequence stars with $0.3 \lesssim
M/\ms \lesssim 1.0$ via magnetically-coupled stellar wind. For some
systems the lifetime to contact is sufficiently shorter than the lifetime of 
helium stars with mass $\sim 0.4$\,\ms. The upper limit of the range of mass of
stars that produce magnetic stellar wind is still not known for certain. 
 If this limit is raised to  1.5\,\ms, the number of
systems with main-sequence companions reduces by 10\%, the discontinuity shifts
to 1.5\,\ms\ and becomes less pronounced.

The low-mass group of white dwarfs  are helium ones, while the group of massive
white dwarfs contains carbon-oxygen and oxygen-neon white dwarfs. Two groups of
white dwarfs overlap on expense of low-mass  ($M \approx 0.35 - 0.40$\,\ms)
hybrid CO-dwarfs; precursors of the latter are helium stars of the same mass.

\subsection{Selection effects}
\label{sec:selection}

In the $\log T_{\rm eff}  - \log g$ diagram helium remnants of stars with thin
hydrogen envelopes spend their whole core-helium burning time in the region
occupied by subdwarfs, since they have $\log T_{\rm eff} = 4.4 - 4.5$ and     
$\log g = 5.65 - 5.90$. For comparison with observations, we have constructed a
subset of model helium stars that is limited by stellar magnitude $V = 16$,
since the main observational surveys of stars with UV-excess are limited by
stellar magnitude $B \approx 16$~\citep{palomar_green}, $B \approx
15.3$~\citep{downes86}. The survey of white dwarfs SPY is limited by $B \approx
16.5$; this catalog also contains a large number of hot subdwarfs due to the
errors of spectral classification in  the input catalog. Since our model is
rather primitive and depends on a number of not very well restricted
parameters, we neglected typical for helium subdwarfs small color excess 
$B-V=-0.25\pm0.1$~\citep{heber86,aznar_jeffery02,stark_wade03,altmann+04}. For
the systems with main-sequence companions we applied, as an effect of
observational selection, the condition $V_{\rm He} < V_{\rm MS}$. Using this
rather rigid condition we get, in essence, the lower limit for the number of
binaries where detection of companion by spectroscopic methods is possible. The
sample limited by stellar magnitude contains only stars with mass lower than
2\,\ms, since for them only it appeared possible to construct the scale of
bolometric corrections for $V$, using computations by \citet{it85}.  Interstellar
absorption was set to $1\m6$/kpc~\citep{allen}. For all binaries we have assumed
that for their detection the semiamplitude of radial velocity has to be
$K_{min} \geq 30$~km/s; this corresponds to the lower limit of $K$ in known
binaries with helium subdwarfs~\citep{morales+03}.

At difference to \citet{hpmmi02,hpmm03}, we did not exclude from
consideration model systems with spectral type G and  K  main-sequence
companions. Han et al. suggested that the objects with Ca II absorption line K 
that were rejected as the candidates to the catalog  PG~\citep{palomar_green},
are, in reality, helium subdwarfs with F-, G-, and K-type main-sequence
companions. However, a special analysis of rejected
candidates~\citep{wade_stark_green04} have shown that they really are
low-metallicity sdF/sdG-subdwarfs, but not sdB/sdO-subdwarfs.

Since we were interested in \textit{all} helium stars, we did not take into
account so called ``strip effect'' -- restriction of the model sample by stars
that have $\log g$ and $\log T_{\rm eff}$ values that are typical for
core-helium burning model helium stars with thin hydrogen envelopes; in fact, some of the stars that are classified as hot subdwarfs are located
outside that ``strip'' (see, e. g.,~\citet{saffer+94,lisker+05}). 

The model sample of stars limited by selection effects contains 15210 objects
(including systems with relativistic companions, see the Table). This number
comprises $\sim 0.4$\% of all helium stars. The catalog PG~\citep{palomar_green}
that is limited by $B \lesssim 16$ contains about 700 objects classified as
helium subdwarfs that are distributed over the area of 10714 sq. deg. This
corresponds to the total number of ``observed'' stars of about 2800. Having in
mind all uncertainties of the model, rather simple account for selection
effects, and misclassification in the PG catalog, we may consider the
agreement in the numbers of stars in our model and in the catalog that is
within factor $\sim 5$ as satisfactory.  The most uncertain parameters of the
model are threshold mass of the progenitors of helium stars and the minimum
mass for helium ignition in the single merger products of white dwarfs (that
constitute about a quarter of total sample). Lowering of both these masses may
improve agreement between the model and observations.

Distribution of model helium stars over mass in the sample  limited by
selection effects is shown in Fig.~\ref{fig:wmall}b.  Reduction of the number
of stars  in the sample as compared to the total sample is caused, in the first
instance, by reduction of the volume of space in which these stars might be
observed (reduction factor $\sim 150-200$). Since we exclude the systems that
have main-sequence stars brighter than their companions, the number of helium
stars with main-sequence companions further reduces by a factor $\sim 20$. As a
result, in the sample limited by selection effects the rate of binarity reduces
to 58\%. Restriction of the sample by the semiamplitude of radial velocity that
is necessary for discovery of binarity practically does not affect the number
of stars in the sample.

Observational estimates of the binary fraction of helium subdwarfs depend on
the sample of stars under study and on the strategy of observations, that
determines the efficiency of detection of binaries. Thus, 
\citet{mhmn01} estimate the binary fraction among sdB stars with periods
from  $0.03$ to $10.0$\,day as  $69\pm9$\%; 
\citet{nap_ehb04}
give  (42 -- 45)\%, depending on assumptions on the real distribution over
periods in the range of $0.03$ to $30.0$\,day and masses of components; 
\citet{saffer+01} estimate the upper limit of the fraction of single
helium stars as 35\%. The reasons for the discrepancy in the observational
estimates of binary fraction among helium subdwarfs may be differences in the
samples of stars that were studied (they could belong to the populations with
different age and metallicity) and the differences in the observational
methods. As it has been shown by 
\citet{hpmm03}, the fraction of
binaries decreases with decrease of metallicity; the sample studied
by \citet{nap_ehb04}, as admit the authors of publication, can contain a
significant admixture of halo and thick disk stars. The model estimate of the 
binary fraction of ``observed'' helium stars agrees with the range of
observational estimates.

Note that reduction of common envelope parameter $\alpha_{\rm ce}$ to 1 results
in increase of mergers in common envelopes; this reduces the fraction of
binaries among helium stars to 35\%, in clear contradiction to observations.

In the total sample of model helium stars the fraction of binaries with
main-sequence components is 38\%; selection effects reduce the fraction of such
systems to 3\%; these binaries may have
composite spectra. The presence of late-type companion may be evidenced by
infrared excess. For instance, according to~\citet{stark_wade03}, the fraction
of systems with unresolved late-type companion among hot subdwarfs may be $\sim
(30 - 40)$\%. This is close to their fraction in the total model sample. In some
cases a main-sequence companion may be detected thanks to the reflection effect,
caused by the heating by hot radiation of the subdwarf. But this effect may be
noticeable only in the most close systems when companion is an M-type star or a
brown dwarf~\citep{maxted_4sdb}. Note however, that for some subdwarfs that have
composite spectra which allow to estimate the parameters of companions, the
masses of the latter appear to be confined to (0.8 -- 1.2)\,\ms
range~\citep{aznar_jeffery02}, i. e. they are close to the maximum of
distribution in the sample limited by selection effects
(Fig.~\ref{fig:wcall}b). The rest of subdwarfs that are classified as single
ones may be really single or have white-dwarf companions.

While the total model population is dominated by helium stars with
main-sequence companions, in the selection effects limited sample systems with
white-dwarf companions start to dominate. The maximum of the distribution of
helium stars over masses in total ``limited'' sample shifts to somewhat higher
masses (into $\log(M/\ms) = -0.40 \div -0.35$ range, see Fig.~\ref{fig:wmall}).
This improves the agreement with the ``canonical'' mass of subdwarfs of
0.5\,\ms.

The distribution of 59 observed sdB/sdO subdwarfs with known orbital periods
(Fig.~\ref{fig:wpsel+obs}a, the data from L. Morales-Rueda, priv. comm.) varies within factor
$\sim 2$ within the interval $0.1 \leq P_{\rm orb} \leq 10.0$\,day. In the
model ``observed'' sample the number of stars varies within the same factor in
the period range $0.3 \leq P_{\rm orb} \leq 3.0$\,day. Relative decrease of the
number of model systems to short periods is comparable to the observed one.
With increase of $P_{\rm orb}$ relative to the maximum of distribution, the
number of stars in the model sample declines much faster than in the observed
sample. There are no stars with $ P_{\rm orb} > 10.0$\,day in the latter
sample. The last circumstance is possibly related to the fact that some stars which
in our model form common envelopes in the first episode of mass exchange, in
reality evolve stably to the longer periods (for instance, due to the loss of a
part of their mass prior to the Roche lobe overflow or due to stabilization of
mass-exchange via momentum loss~\citep{hte00}. This effect requires a test by
computation of grids of evolutionary models.

Accumulation of long series of observations will, most probably, result in
discovery of new subdwarfs with long orbital periods. However, it is necessary
to have in mind that the efficiency of binarity detection declines with
increase of $P_{\rm orb}$: it exceeds   80\% at  $P_{\rm orb} \leq 10.0$\,day,
decreases to 50\% at  $P_{\rm orb} \sim 20.0 -
30.0$\,day~\citep{mhmn01,nap_ehb04},   and then rapidly declines to 0. One would
expect that observations with high spectral resolution would result in
detection of subdwarfs with relatively long orbital periods. But, for instance,
among nine stars detected in SPY survey, seven have $P \lesssim 1.0$\,day, but only two
-- $P_{\rm orb} = 5.87$ and  $7.45$\,day~\citep{nap_ehb04}.

\subsection{Relations between parameters of binaries with nondegenerate helium components}
\label{sec:relations}

\begin{figure}[!t]
\includegraphics[scale=0.66,angle=-90]{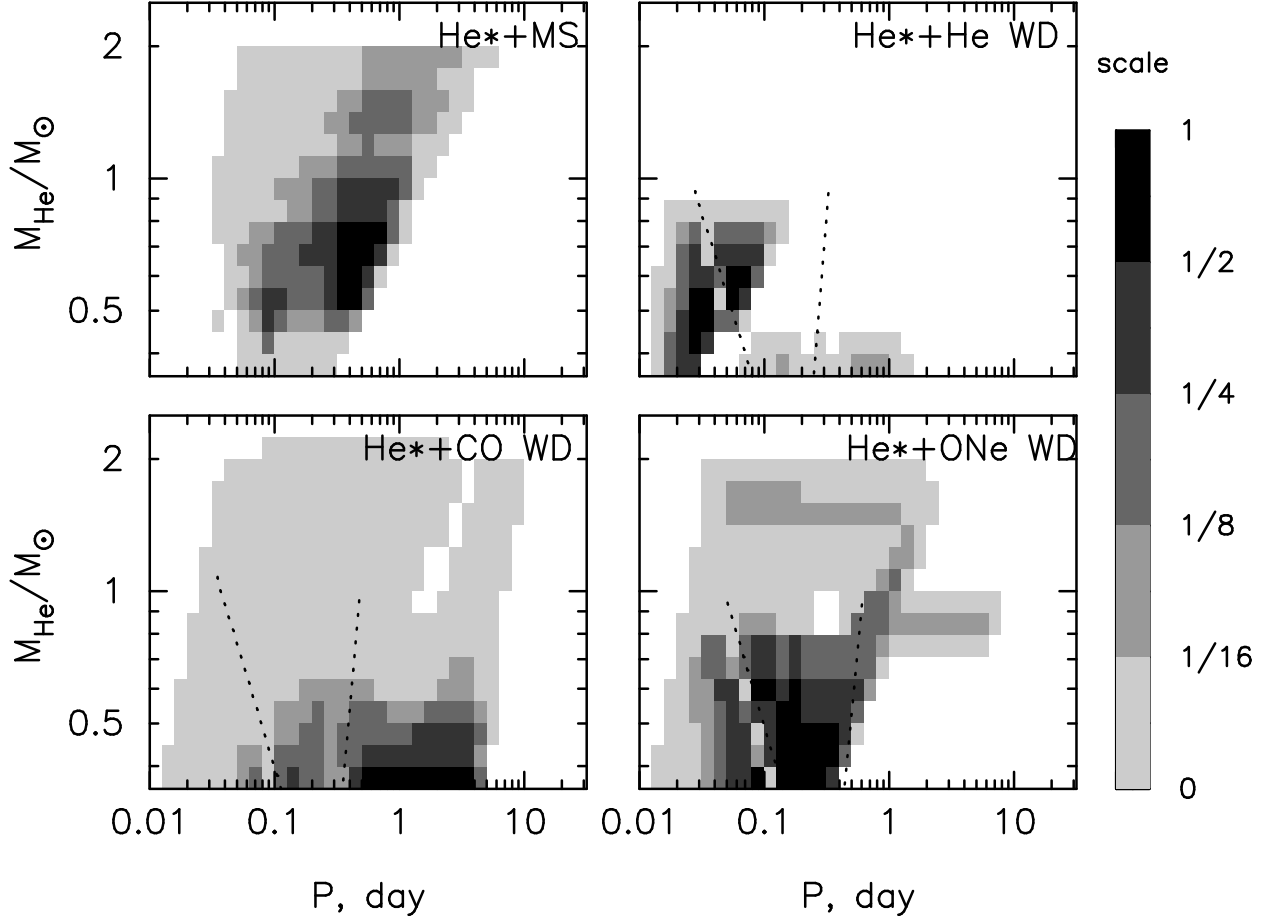}
\caption{Relations between orbital periods and masses of helium stars -- components
of  binary systems  that have secondaries of different types. The sample
limited by selection effects is shown. Five grades of gray-scale show the
regions where the number density of systems is within 1/2 of the maximum of
${{{\partial^2{N}}\over{\partial {\log P}}{\partial {\log M}}}}$, or is confined
to the ranges of 1/4 -- 1/2, 1/8 -- 1/4, 1/16 -- 1/8, and 0 -- 1/16 of the
maximum. Each panel is scaled separately. In the blank regions of the diagram systems with white dwarfs are
absent. In the panels for binaries with white-dwarf companions, the systems
that have  at birth orbital periods shorter than those marked by the left
dotted line merge in the lifetime of helium stars. Systems that have periods at
birth shorter than the ones marked by right dotted line merge in less than
Hubble time.}
\label{fig:pm1}
\end{figure}

\begin{figure}[t!]
\hskip -1 cm
\parbox{\textwidth}{\includegraphics[scale=0.8]{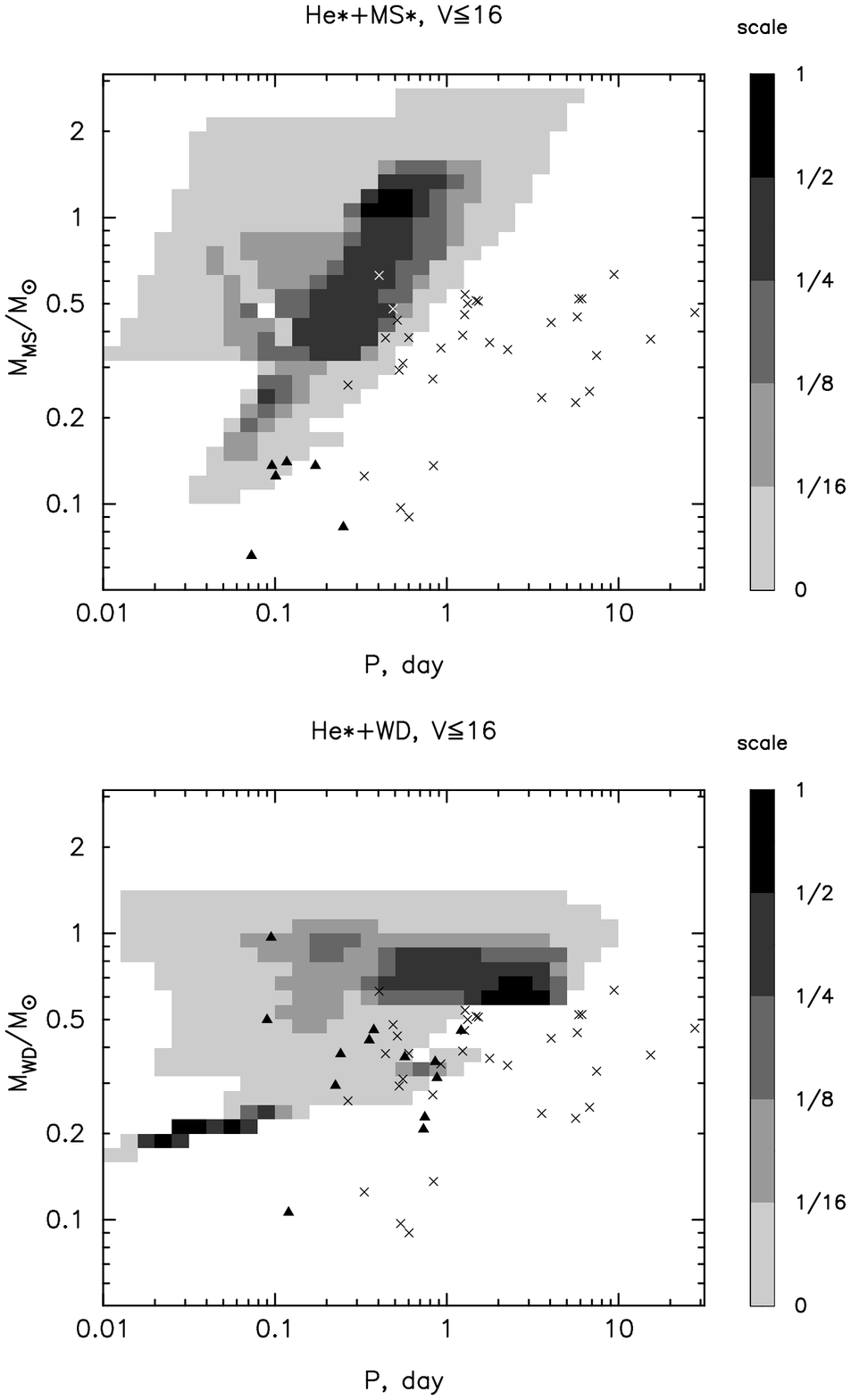}}
\vskip -2cm
\caption{Relation between orbital periods of binaries and masses of companions to
helium stars -- main-sequence stars and white dwarfs. The gray-scale is similar
to that in Fig.~\ref{fig:pm1}. In the upper panel triangles show known masses
of main-sequence companions or their lower limits, in the lower panel triangles
show masses or lower limits to mass of known white-dwarf companions. In both
panels the crosses show lower limits to the mass of unknown type companions.  
}
\label{fig:pm2}
\end{figure}

In Fig.~\ref{fig:pm1} we show the number
distributions of binaries with helium components and different companions in
the selection limited model sample in the ``logarithm of the orbital period
$\log P_{\rm orb}$ -- logarithm of helium star mass $\log M_{\rm He}$''
diagram.   As we noted before, after taking observational selection into
account, among ``observed'' helium-star binaries dominate the systems with
companions -- white dwarfs ($\sim 90$\%). In these binaries the masses of
helium stars have to be confined between minimum mass of 0.35\,\ms\ and 
0.5\,\ms. In scarce systems with main-sequence components somewhat more massive
helium stars may be observed, with masses up to (0.8 -- 1.0)\,\ms.

Distributions of systems with different white dwarfs over periods differ: at 
$P \lesssim 1.0$\,day~ it is more probable to expect the presence of a helium
white dwarf than that of a carbon-oxygen or an oxygen-neon one. Despite we know
from observations only minima of the masses of white dwarf companions to helium
stars, let note that only in 4 out of 12 known sdB+WD systems with  $P_{\rm
orb} \leq 1.0$\,day we may exclude the presence of helium white dwarfs
($M_{\rm WD}^{min} \geq 0.4$\,\ms). In two known systems with $P_{\rm orb} >
1.0$\,day white dwarfs have to belong to carbon-oxygen/oxygen-neon family (see
Table~5 in ~\citep{morales+03}).

In Fig.~ \ref{fig:pm2}  we show, for the same model
sample limited by selection effects, the distributions of systems with helium
stars and main-sequence or white-dwarf companions in the ``logarithm of the
orbital period $\log P_{\rm orb}$ -- logarithm of companion mass $\log M_2$'' 
diagram. We plot in the Figure the masses of companions to helium stars known
from observations, as well as the estimates of the lower limits to the masses
of companions, if only mass function $f_m$ is known: 
$$f_m = \frac{M_2^3 \sin^3
i} {(M_1+ M_2)^2} = \frac{P_{\rm orb} K_1^3} {2 \pi G}. $$ 
We use the data on
mass, semiamplitude of radial velocity ($K_1$) and orbital period for 52
objects listed in~\citep{morales+03,nap_ehb04} and unpublished data provided to
us by L. Morales-Rueda. For the estimate of $M_2^{min}$  we assumed that
$M_{\rm He}$=0.5\,\ms, $\sin i=1$. Assumption of  $M_{\rm He} = 0.35\,M_\odot$\
only weakly influences location of stars in Fig.~\ref{fig:pm2}. Variation of
orbital inclination from 90$^\circ$ to the most probable one  $\sim 60^\circ$
increases $M_2$ by a factor close to 2.  In order to shift all observed objects
in Fig.~\ref{fig:pm2} in the region populated by model systems, it is
necessary, as a rule,  to assign to observed systems quite probable orbital
inclination angles of  20$^\circ$ -- 30$^\circ$.

Several binary sdB stars with M-type companions were discovered due to the eclipses
or reflection effect in the most close systems with $P\lesssim 0.25$\,day. As
we noted above, companions with mass of  $1 \pm 0.2\,M_\odot$~ are suspected in
some systems with composite spectra, but among systems with known orbital
periods they are absent.

Since the number of model stars with white-dwarf components is by a factor
close to 200 higher than the number of systems with main-sequence companions,
one may expect that the overwhelming majority of the new identified and 
discovered companions to helium subdwarfs will be carbon-oxygen or helium white
dwarfs. As we mentioned above, the deficit of long-period model binaries may be
related to the underestimate of the fraction of binaries with the stable first mass
exchange.

In the subsample of helium stars limited by observational selection, about 30\%
of stars are single. The mass spectrum of these stars is slightly shifted to
larger periods, as compared to the spectrum of binary subdwarfs
(Fig.~\ref{fig:wmall}). These stars may be probably related to the sdВ/sdO
subdwarfs that have atmospheres enriched in He and that are slightly hotter
than sdB subdwarfs with  hydrogen-rich atmospheres. Location of these stars in
the $\log T_{\rm eff} - \log g$ diagram is more similar to  the tracks of
merger products of helium dwarfs than to the tracks of single low-mass stars
with helium cores and thin hydrogen envelopes~\citep{ahmad_jeffery03}. Let note,
that the only sample of 23 sdO-stars that was systematically explored for
binarity, contains only one binary star~\citep{nap_ehb04}! Special study of
binarity detection probability by \citet{nap_ehb04} have shown, that the binary
fraction in this sample is really close to only 5\%.

\subsection{Final stages of the evolution of close binary stars with low-mass helium components}
\label{sec:final}

As we noted above, in the closest systems with helium components the latter may
fill their Roche-lobes before helium exhaustion in their core, because of the loss
of systemic angular momentum via gravitational wave radiation. Respective
limits of orbital periods are plotted in Fig.~\ref{fig:pm1} for the typical
mass of white dwarfs $M_{\rm He}=0.2$\,\ms,   $M_{\rm CO}=0.6$\,\ms,     
$M_{\rm ONe}=1.3$\,\ms. If conditions for stable mass loss are fulfilled,
AM~CVn systems arise~\citep{ty96,npv+01}. The birthrate of such systems in our
model is $\simeq 5 \times 10^{-4}$~yr$^{-1}$. Rather similar estimate ($\simeq
3 \times 10^{-4}$~yr$^{-1}$) was obtained in~\citet{nyp04} for somewhat
different assumptions on star formation history and stellar evolution, but with
the same Eq.~(\ref{eq:j}) for description of mass exchange in the systems of
main-sequence stars with comparable masses of components. Evolution of AM~CVn
stars and their observed features are considered in detail
by \citet{ty96,npv+01,nyp04}. If conditions for stable mass exchange are not
fulfilled, the merger of helium stars with carbon-oxygen or oxygen-neon white
dwarfs possibly results in formation of R~CrB stars~\citep{ity96}.

After helium exhaustion in their cores, helium stars with mass $\lesssim
0.8$\,\ms\ turn into carbon-oxygen or oxygen-neon white dwarfs of the same mass
almost without any expansion. In this way close  binary white dwarfs are formed
(another scenario leads to formation of close binary white dwarfs through case
C of mass exchange). If binary white dwarfs have short enough orbital periods,
they may merge due to the angular momentum loss via radiation of gravitational
waves in less than Hubble time. Respective limiting periods are shown in
Fig.~\ref{fig:pm1}. Currently (March 2005) the data on orbital periods of 19
close binary white dwarfs is published. Six of these systems (see, e. g.,
~\citet{nap_dubr} and Table 11 in~\citet{morales_6}) have periods short enough
for merger in Hubble time and one of these systems has total mass larger than
the Chandrasekhar one. If both of components of this system are carbon-oxygen
white dwarfs, it may occur to be a precursor of type Ia
supernova~\citep{ty81,it84a,web84}. The merger of  carbon-oxygen and helium
white dwarfs, most probably, results not in SN~Ia, but in ``helium
Novae''~\citep{efy01,yoon_langer04} or in formation of R~CrB stars~\citep{ity96}.
Note, that our model predicts the rate of merger of carbon-oxygen white dwarfs
with super-Chandrasekhar total mass equal to 0.0013 per yr, slightly lower than
the prediction for the Galactic rate of SN~Ia equal to $\nu_{\rm
SNIa}=(0.004\pm0.002)$~per yr~\citep{cet99,cap01} (based on the assumption that Milky Way
has  $\nu_{\rm SNIa}$ equal to the average one for the galaxies of the same
morphological type Sb-Sbc and blue luminosity
$L_B=2.3\times10^{10}L_{B,\odot}$). Theoretical and observational estimates of 
$\nu_{\rm SNIa}$ may be brought to a better agreement by reduction of
$\alpha_{ce}$ in Eq.~(\ref{eq:ce})~\citep{ty02}. Of course, it is quite possible
that the merger of white dwarfs is not the only mechanism for SNe~Ia (see, e.
g.,~\citet{yu05}); the numerical relation between different scenarios of
evolution that may potentially lead to SNe~Ia is still a function of numerous
parameters of the population synthesis. Note however, that our trial computations
have shown that reduction of $\alpha_{ce}$ shifts the maximum of the distribution
of double helium stars to too low, as compared to observations, orbital periods
(Fig.~\ref{fig:wpsel+obs}). The situation will become more clear when
sufficient data on close binary helium stars and white dwarfs will be
accumulated.

\section{Conclusion}
\label{sec:concl}

We have modeled population of low-mass helium stars in the Galaxy under assumption that all of them are formed in close binary stars. We got estimates of their birthrate  --
$0.043$\,yr$^{-1}$, total number --  $4 \times 10^6$, and binary fraction -- 76\%. These estimates are in a good agreement with our analytical estimates~\citep{ty90} and with estimates by
\citet{hpmm03} that are also obtained by means of population synthesis.

Low-mass helium stars are identified with observed sdB/sdO subdwarfs. The study
of observational selection effects shows that the dominant factor in the formation
of the observed ensemble of sdB/sdO stars is the fact that observed samples are
limited by relatively bright stellar magnitude ($V\lesssim 16$).The number of
helium stars with main-sequence companions available for observations is
additionally limited by the circumstance that majority of subdwarfs  are by far
brighter than their companions. All this  reduces the binarity rate in the
model observed sample to 58\%; the latter value agrees with observations.
According to our computations, in the observed population of binary helium
subdwarfs overwhelming majority of companions to helium stars are carbon-oxygen
or helium white dwarfs.

In our model, single low-mass helium stars are, predominantly, merger products
of helium white dwarfs. Since merge mainly low-mass dwarfs, single helium
subdwarfs have masses close to those of components of binaries. It is possible,
that a fraction of merger products may be identified with sdO subdwarfs that,
according to existing rather scarce data, have a very low binarity rate.

It is assumed in our model that all stars are formed in binaries, but despite
this we are able  to explain, quite naturally, the formation of helium stars by
the evolution in \textit{close} binaries. As well, we can  explain  the
binarity rate of low-mass helium stars. On the other hand,  
\citet{saffer+01}, based on observations of radial velocities of subdwarfs,
separate them into three groups. In the spectra of stars belonging to the first
group the features of companions are absent and radial velocities of these
stars are virtually constant. In the spectra of stars in the second group the
features of companions are also absent, but variability of radial velocities
indicates orbital periods  $\sim 1.0$\,day. In the spectra of the third group
stars the features of cold companions are present. The percentage of stars
belonging to these three groups is  35, 45, and  20\%. The following
phenomenological interpretation is suggested~\citep{saffer+01,green+01}. The
stars of the first group are intrinsically single objects that lost a part of
their mass not long before helium ignition in their cores (see scenario
suggested by \citet{dcruz96}). The stars in the second group are binaries that
passed through the common envelope and have invisible companions -- M-type
stars or white dwarfs. In the third group there are stars that passed through
the stage of stable mass exchange in the red giants stage. If the model
suggested by \citet{saffer+01,green+01}  will  get additional confirmation, for
instance, distributions of stars over orbital periods will be found, companion
masses will be estimated, selection effects that define the ratio between
different groups of stars will be established, then our model for the
population of low-mass helium stars and the analysis of selection effects that
form observed sample of stars would require certain revision.

\vskip 0.5 cm

The authors acknowledge L. Morales-Rueda for providing a compilation of known orbital periods of subdwarfs and unpublished data. We are indebted to R. Napiwotzki, G. Nelemans,
for useful remarks
 and to E. Green for providing  electronic files of papers otherwise unavailable in Russia. This study was partially supported by Russian Foundation for Basic Research grant (project code 03-02-16254), the program of the Section of General Physics and Astronomy of the Russian Academy of Sciences ``Extended Objects in the Universe'', ``Leading Russian Science Schools Support Program'' (project code
НШ-162.2003.2), and by the grant ``Non-stationary Processes in  the Universe''.


\end{document}